\documentclass[]{interact}

\usepackage{epstopdf}
\usepackage[caption=false]{subfig}

\usepackage[numbers,sort&compress,merge]{natbib}
\bibpunct[, ]{[}{]}{,}{n}{,}{,}

\theoremstyle{plain}

\theoremstyle{definition}

\theoremstyle{remark}

\usepackage{hyperref}
\usepackage{xcolor}

\newcommand{\pa}[1]{\textcolor{black}{#1}}

\newcommand{\Pe}{\text{Pe}}

\begin{document}


\title{Coil-to-globule collapse of active polymers: a Rouse perspective}

\author{
\name{Paolo Malgaretti\textsuperscript{a}\thanks{CONTACT P. Malgaretti. Email: p.malgaretti@fz-juelich.de},  Emanuele Locatelli\textsuperscript{b,c}, and Chantal Valeriani\textsuperscript{d,e}}
\affil{\textsuperscript{a} Helmholtz Institute Erlangen-N\"urnberg for Renewable Energy (IET-2), Forschungszentrum J\"ulich, Erlangen, Germany}
\affil{\textsuperscript{b} Department of Physics and Astronomy,  University  of  Padova, 35131 Padova,  Italy}
\affil{\textsuperscript{c}INFN, Sezione di Padova, via Marzolo 8, I-35131 Padova, Italy}
\affil{\textsuperscript{d} Dep. Est. de la Materia, F\'isica T\'ermica y Electr\'onica, Universidad Complutense de Madrid, 28040 Madrid, Spain}
\affil{\textsuperscript{e}
GISC - Grupo Interdisciplinar de Sistemas Complejos 28040 Madrid, Spain
}
}

\maketitle

\begin{abstract}
We derive an effective Rouse model for tangentially active polymers, characterized by a constant active force tangent to their backbone. In particular, we show that, once extended to account for finite bending rigidity, such active Rouse model captures the reduction in the  gyration radius, or coil-to-globule-like transition, that has been observed numerically in the literature for such active filaments. Interestingly, our analysis identifies the proper definition of the Peclet number, that allows to collapse all numerical data onto a master curve.\vspace{0.2cm}
\\\resizebox{25pc}{!}{\includegraphics{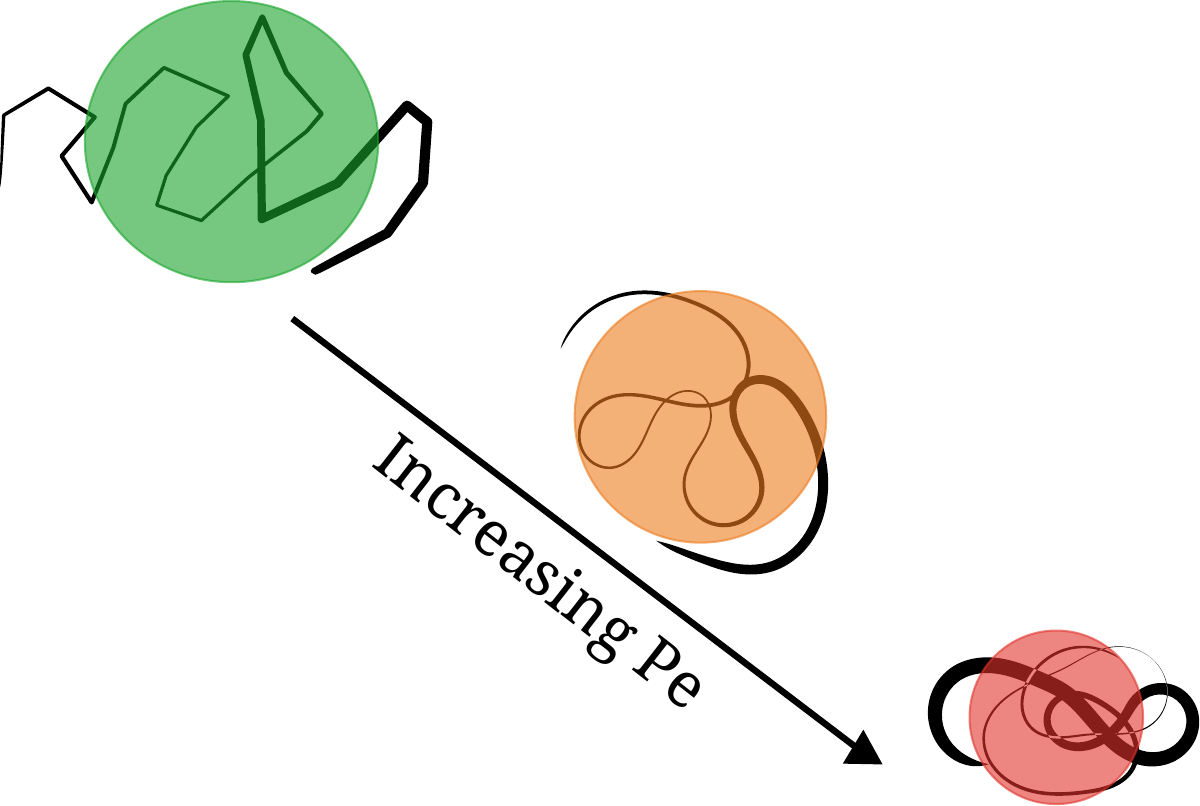}}

\end{abstract}


\section{Introduction}
By means of breaking equilibrium at the {local, microscopic} scale, active systems show  {dynamical and collective properties} that differ quite much from their equilibrium or even driven counterparts\cite{marchetti2013hydrodynamics, fodor2018statistical}. For example,  {a collection of active Brownian colloids} can undergo Motility Induced Phase Separation\cite{cates2015motility,schwarz2012phase} leading to the onset of big clusters even in the absence of any attractive interaction between the particles. Similarly, micro-phase separation has been observed in continuum models that mimic an active bath\cite{wittkowski2014scalar}. 
Besides MIPS, active systems present out-of-equilibrium phases such as living crystalline clusters \cite{Palacci2013,mognetti2013}, active turbulence \cite{Cisneros2010},  self-assembly \cite{Murugan2015,Mallory2018} and various types of flocking phases \cite{Narayan2007,Hayakawa2020,Suematsu2010}. \\
\noindent
{Among active systems, those made of filamentous units have particular relevance in biological systems, example being the cytoskeleton\cite{fletcher2010cell} and the intracellular trafficking network\cite{vale2003molecular}, chromatin\cite{mahajan2022euchromatin, goychuk2023polymer, shin2024transcription}, cilia arrays\cite{loiseau2020active, chakrabarti2022multiscale} and flagella\cite{chelakkot2014flagellar} as well as micro-organisms\cite{faluweki2023active, patra2022collective, rosko2024cellular}. At the macroscopic scale, worms collectives show interesting emerging properties\cite{deblais2023worm}. More generally, technological progress in the synthesis of artificial active chains\cite{Dreyfus2005a, Hill2014, Biswas2017, Nishiguchi2018} as well as chains of chemically active droplets\cite{kumar2023emergent, subramaniam2024emergent} and soft robotic systems\cite{ozkan2021collective, savoie2023amorphous, becker2022active} make active filaments ubiquitous and open  up the possibility of a huge range of applications.\\ 
Inspired by these examples, focus} has been recently put onto  {characterising the properties} of active polymers, i.e. polymers made out of ``active'' monomers.  {In this context, activity can be realised in different ways\cite{Winkler2020}: by means of a temperature mismatch\cite{ganai2014,Smrek2017, Active_topoglass_NatComm20}, correlated noise along the backbone\cite{osmanovic2017dynamics,  goychuk2023polymer}, completely random self-propulsion forces (or Active Brownian Polymer)\cite{Kaiser2014, eisenstecken2016conformational, Das2019, theeyancheri2024dynamic} or correlated forces, oriented either perpendicularly\cite{prathyusha2022emergent} or along the polymer backbone\cite{Isele-Holder2015, bianco2018globulelike}. In this manuscript, we focus on the last case, as it is believed to mimic the action of molecular motors \pa{under suitable conditions ~\cite{Terakawa2017,vliegenthart2020filamentous} (see Ref.~\cite{Grosberg2023} for counter-examples)}  as well as the locomotive mechanism of worms~\cite{nguyen2021emergent, deblais2023worm}, that contract their segments or use lateral protrusions to crawl or swim forward.\\
Notably, the way in which a tangential force can be realized is not unique. Indeed, one can choose to consider a propulsion force (i) constant in magnitude and parallel to the local backbone tangent\cite{bianco2018globulelike, mokhtari2019dynamics,das2019dynamics, foglino2019,Locatelli2021,patra2022collective,li2023nonequilibrium,miranda2023self, vatin2024conformation,Lamura2024}; (ii) constant in magnitude and parallel to the bond between neighbours along the chain\cite{Singh2018, prathyusha2018dynamically, abbaspour2023effects, kurzthaler2021geometric}; (iii) proportional to the bond vector\cite{Isele-Holder2015, peterson2020statistical, philipps2022tangentially, fazelzadeh2022effects, Farouji2023}. Notably, these slightly different definitions leads to some discrepancies in the steady state conformations. In particular, in case (i) a globule-like transition has been reported in three dimensions\cite{bianco2018globulelike,li2023nonequilibrium}, where polymers assume more and more compact conformations upon increasing the strength of the propulsion. So far, a theoretical explanation of this phenomenon is lacking.\\
In this manuscript, we propose a minimal model of an active polymer that displays such globule-like transition, and  solve it by means of a hybrid analytical/numerical approach. The proposed model recapitulates the central role of the choice of a constant backbone propulsion and offers a way to propose a new definition of the activity, the so-called P\'eclet number. In what follows, we will introduce and develop the model more in detail. \pa{We stress that the derivation of the model does not rely on a systematic expansion, rather we have tried to identify the "core" that is necessary to reproduce the phenomenology that has been reported by numerical simulations. In doing so, we make assumption that not always can be casted in a formal expansion. That is why, rather then discussing the details of the approximations that we do, or the "order" of the terms that we disregard, we assess the validity of the proposed model by comparing its predictions against the numerical results available in the literature. Interestingly, our model provides a remarkably good  quantitative agreement with the numerical data hence validating, a posteriori, the assumptions that we made to derive it. In particular,} we will show that a minimal continuous model can be defined only if a finite bending rigidity is included. 
We will further discuss consequences of the model, such as the emergence of a master curve onto which data for filaments of different length and activity collapse.}  

\section{Model}
\label{sec:mod}
\begin{figure}
    \centering
    \includegraphics[width=0.65\textwidth]{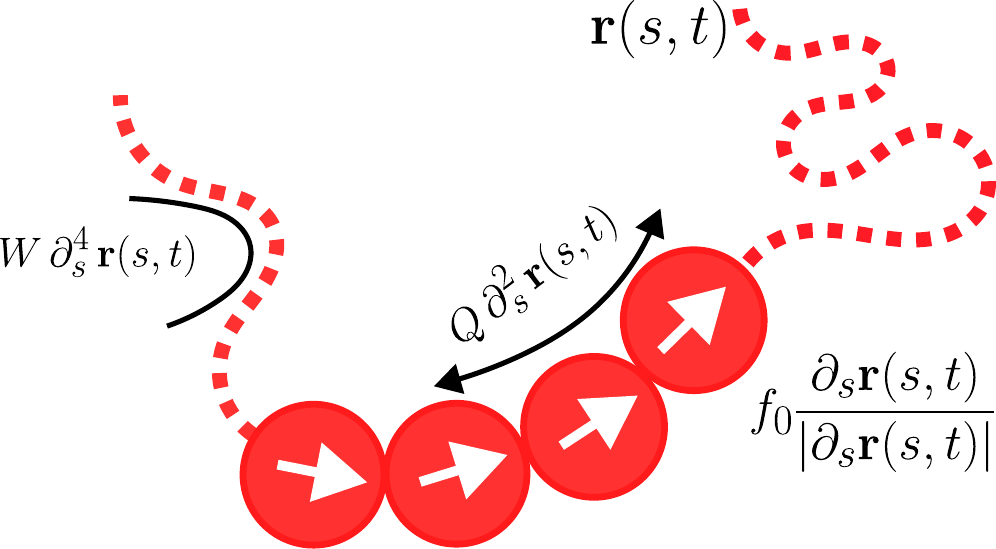}
    \caption{Schematic representation of an active polymer as a succession of active beads. The self-propulsion direction, for each monomer, has constant magnitude and is directed as the backbone tangent vector.}
    \label{fig:cartoon}
\end{figure}

We consider a tangentially active polymer, i.e. the self-propulsive force of each monomer acts along the  backbone's tangent. In particular, we consider the case (i), as named in the Introduction, where the propulsion force is constant in magnitude and parallel to the local backbone tangent. The system is considered in the overdamped limit. To exemplify, let's imagine a necklace of active colloids, such as diffusiophoretic Janus particles\footnote{\pa{In this example we assume that the diffusiophoretic Janus particles are in the reaction-limited regime within which diffusion is so fast as compared to the reaction rate that density gradients can be neglected and hence the "phoretic" interactions~\cite{Liebchen2019} between colloids can be disregarded.}}~\cite{Liebchen2019,theurkauff2012dynamic}, joined in such a way that the "south" pole of a colloid is {, at all times,} in contact to the "north" pole of the other one (see Fig.~\ref{fig:cartoon}). 
While the system composed by the polymer and the solvent is force-free, as the active colloids are ``swimmers'', there is a net force on each monomer composing the backbone of the polymer.\\
{As, according to our construction, the orientations of the beads are constrained, such a force is bound to} act along the tangent direction to the backbone.  {Further, disregarding the} depletion of reactants or more complex interactions, the magnitude of the force can be regarded as fixed and independent of the polymer configuration, {as it is a characteristic of the monomeric unit.} 
The discretized, bead-spring realisation of such an active polymer model in three dimensions has been investigated numerically in  Refs.~\cite{bianco2018globulelike,Locatelli2021,li2023nonequilibrium,miranda2023self,vatin2024conformation}\\
We remark that these results are qualitatively different from theoretical calculations\cite{peterson2020statistical, philipps2022tangentially} and are, for self-avoiding polymers, slightly yet appreciably different from those reported in Refs.~\cite{Singh2018,fazelzadeh2022effects}. As already suggested in the literature, in cases (ii) and (iii) the magnitude of the active force on each monomer is not strictly homogeneous and it varies within a range. Indeed, the propulsion force is split between neighbours and each monomer (except the ones at the extremes of the polymer) gets a contribution from each bond. The resulting force is proportional to the tangent vector: as such, bent conformations experience a smaller propulsion than straight ones. This difference is probably at the heart of this discrepancy, as will be  highlighted by the minimal model tackled in this paper. As such, despite the seemingly formal difference, these qualitative discrepancies may be sufficient to identify case (i) as a different model from (ii) and (iii), at least in three dimensions. {From a modeling perspective, one may argue that case (i) could be more suitable if monomers generate their own propulsion or if they are individually pushed by some external agent, such as a molecular motor. Instead, case (ii) and (iii) may be more suitable for coarse-grained representations, where monomers are effective units and the active force may result from the streaming of  motors; alternatively, one may consider  systems where molecular motors are strong enough to push more than one monomer.}

\section{Building a continuous minimal active polymer model}

In what follows, we will introduce a continuous minimal model for a tangentially active polymer. The polymer is described as a curve $\mathbf{r}(s,t)$, parameterized by the dimensionless contour position $s\in[-N/2,N/2]$ that moves along the polymer backbone, which is subject to active forces, constant in module and related to the polymer conformation. The filament is also subject to random, thermal noise $\boldsymbol{\eta}(s,t)$, that satisfies the usual fluctuation-dissipation relations
\begin{equation}
\left\langle \boldsymbol{\eta}(s,t)\right\rangle   =0 \qquad \left\langle \boldsymbol{\eta}_{i}(s,t)\boldsymbol{\eta}_{j}(s',t')\right\rangle   =2\mu k_{B}T\delta_{ij}\delta(t-t')\delta(s-s').
\label{eq:eta}
\end{equation}

\subsection{Fourier representation}

As we will perform most of our analytical calculations in Fourier space, we introduce  the standard decomposition in planar waves:
\begin{equation*}
\mathbf{r}(s,t) =\frac{1}{\sqrt{N}}\sum_{n=-\infty}^{\infty}\mathbf{r}_{n}(t)e^{\imath k_{n}s}\label{eq:r_fourier}\qquad
\boldsymbol{\eta}(s,t) =\frac{1}{\sqrt{N}}\sum_{n=-\infty}^{\infty}\boldsymbol{\eta}_{n}(t)e^{\imath k_{n}s}
\end{equation*}
with $k_n = \frac{\pi n}{N}$. We recall that the amplitudes are defined as 
\begin{align}
    \mathbf{r}_n(t) = \int_{-\frac{N}{2}}^{\frac{N}{2}}\mathbf{r}(s,t)e^{-\imath k_n s} ds
\end{align}
and that the chosen basis is not orthonormal
\begin{align}
    \int_{-\frac{N}{2}}^{\frac{N}{2}} e^{\imath (k_n- k_m)s}ds= 2\dfrac{\sin(\frac{N}{2}(k_m-k_n))}{k_m-k_n}
\end{align}
For $\boldsymbol{\eta}_n$ it holds 
\begin{equation}
\left\langle \boldsymbol{\eta}_{n}(t)\right\rangle =  0 \qquad \left\langle \boldsymbol{\eta}_{n,i}(t)\boldsymbol{\eta}_{m,j}(t')\right\rangle = 2\mu k_{B}T\delta_{ij}\delta_{n,m}\delta(t-t')
\label{eq:noise_ampl}
\end{equation}
In order to enforce the reality of $\mathbf{r}(s,t)$ and $\eta(s,t)$ we have $\mathbf{r}_{n} = \mathbf{r}_{-n}^{*}$, $\boldsymbol{\eta}_{n} =  \boldsymbol{\eta}_{-n}^{*}$. 

\subsection{Active Rouse Model}

{In the continuum limit}, we model the active polymer by adding the constant tangential force to the usual Rouse model 
\begin{equation}
\dot{\mathbf{r}}(s,t)=
\mu Q\partial_{s}^{2}\mathbf{r}(s,t)+\mu f_0 \frac{\partial_{s}\mathbf{r}(s,t)}{|\partial_{s}\mathbf{r}(s,t)|}+\boldsymbol{\eta}(s,t)
\label{eq:Rouse}
\end{equation}
$\mu$ is the mobility of the monomers with length $b$, $f_0$ is the active force, 
$Q$ is the strength of the monomer-monomer interactions. 
We remark that Eq.~\eqref{eq:Rouse} should be completed with a set of boundary conditions. For the case of free ends, i.e. no forces on the head and tail of the polymer, the boundary conditions read~\cite{Winkler1995,Winkler2016}: 
\begin{align}
 2Q\partial_s\mathbf{r}(s,t)\Big|_{s=\pm N/2} = 0
 \label{eq:BC_model-Rouse}
\end{align}
In order to get analytical insights of Eq.~\eqref{eq:Rouse},  one typically looks for the eigenfunctions of the operator on the rhs of Eq.~\eqref{eq:Rouse} that are also compatible with the boundary conditions, Eq.~\eqref{eq:BC_model-Rouse}. For the case under study, this is a formidable task due to the non-linear forcing term. In order to avoid this difficulty we propose a strong assumption and avoid imposing the boundary conditions summarised in Eq.~(\ref{eq:BC_model-Rouse}). This amounts to introducing forces and torques on the edges of the polymer whose magnitude, direction and time correlation are, in principle, out of control and can be determined a posteriori. \pa{We will discuss this issue again once we introduce the Fourier representation}.
Accordingly, Eq.(\ref{eq:Rouse}), in its Fourier representation, reads
\begin{align}
\sum_{n=-\infty}^{\infty}\dot{\mathbf{r}}_{n}(t)e^{\imath k_{n}s}=&-\mu \sum_{n=-\infty}^{\infty}k_{n}^{2}Q\mathbf{r}_{n}(t)e^{\imath k_{n}s}\nonumber\\
&+\dfrac{\sum_{n=-\infty}^{\infty}   \imath \mu f_0 k_{n}\mathbf{r}_{n}(t)e^{\imath k_{n}s}}{\sqrt{\sum_{m=-\infty}^{\infty} k_m \mathbf{r}_{m}e^{\imath k_{m}s} \cdot\sum_{n=-\infty}^{\infty} k_n \mathbf{r}^*_{n}e^{-\imath k_{n}s}}}+\sum_{n=-\infty}^{\infty}\boldsymbol{\eta}_{n}(t)e^{\imath k_{n}s}
\end{align}
by multiplying both sides by $e^{-\imath k_{j}s}/N$, using the sum rule (reported in Eq.~\eqref{eq:sum_rule}) and  {performing the integral}
in $ds$, we obtain:
\begin{align}\label{eq:mode_NL_0}
\sum_n &\frac{\sin\left(\frac{N}{2}(k_j-k_n)\right)}{N(k_j-k_n)}\dot{\mathbf{r}}_{j}(t)=-\mu \sum_n k_{n}^{2}Q\mathbf{r}_{n}(t)\frac{\sin\left(\frac{N}{2}(k_j-k_n)\right)}{N(k_j-k_n)}\nonumber\\
&+\frac{\imath \mu f_0}{N}\int \dfrac{\sum_{n=-\infty}^{\infty}  k_{n}\mathbf{r}_{n}(t)e^{\imath k_{n}s}}{\sqrt{\sum_{m=-\infty}^{\infty}k_m \mathbf{r}_{m}e^{\imath k_{m}s}\cdot \sum_{n=-\infty}^{\infty} k_n \mathbf{r}^*_{n}e^{-\imath k_{n}s}}} e^{-\imath k_{j}s} ds+\boldsymbol{\eta}_{j}(t)
\end{align}
\pa{Coming back to the boundary condition, Eq.~\eqref{eq:BC_model-Rouse}, in the Fourier representation it reads
\begin{align}
    \sum_n k_n \mathbf{r}_n(t) e^{\pm\imath k_n N/2} =0
\end{align}
and it implies a coupling between the amplitudes of the modes. In order to solve these equations numerically we need to truncate the number of modes at $n=N_{cut}$ and hence the boundary condition can be "absorbed" in the truncation process. Since this procedure is not formally correct one should check a posteriori that the effective forces and torques induced by such a choice are small enough to be neglected.}
In order to exploit the Fourier analysis we need to explicitly calculate  the integral on the rhs of Eq.~\eqref{eq:mode_NL_0}. In order to do so we rewrite the denominator as
\begin{align}
  &\sum_{m=-\infty}^{\infty}k_m \mathbf{r}_{m}e^{\imath k_{m}s}\cdot  \sum_{n=-\infty}^{\infty} k_n \mathbf{r}_{n}e^{\imath k_{n}s}=\sum_{n=-\infty}^{\infty}k_n^2 \mathbf{r}_{n}\cdot\mathbf{r}_n^*-\sum_{n=-\infty}^{\infty}\sum_{m\neq -n}k_m k_n \mathbf{r}_{m}\cdot\mathbf{r}_{n}e^{\imath (k_{m}+k_{n})s}
  \label{eq:denom}
\end{align}
We remark that at equilibrium, $f_0=0$, and we have 
\begin{align}
    \langle \mathbf{r}_{n,i} \mathbf{r}_{n,j}\rangle_0 = \delta_{nm}\delta_{ij} \frac{b^2 N^2}{3\pi^2}\frac{1}{n^2}
    \label{eq:equil}
\end{align}
Our approach here is to expand the amplitudes around equilibrium \pa{for small values of the active force, $\frac{f_0 b}{k_BT}\ll 1$. Accordingy, the amplitude of the correlators among Fourier modes can be expressed as 
\begin{align}
   \langle \mathbf{r}_{n,i} \mathbf{r}_{n,j}\rangle \simeq \langle \mathbf{r}_{n,i} \mathbf{r}_{n,j}\rangle_0 +  \frac{f_0 b}{k_BT}\langle \mathbf{r}_{n,i} \mathbf{r}_{n,j}\rangle_1 + \mathcal{O}\left(\frac{f_0 b}{k_BT}\right)^2
   \label{eq:rr-approx}
\end{align}
Accordingly, substituing Eq.~\eqref{eq:rr-approx} into Eq.~\eqref{eq:denom}, and recalling that $k_n=\frac{\pi n}{N}$,} we observe that the sum of the first term in the rhs of Eq.~\eqref{eq:denom} is diverging in the limit $n\rightarrow \infty$ whereas the second term remains finite. This implies that  {the contribution of the active force is always null and} the proposed Rouse model cannot capture the coil-to-globule transition observed in the numerical simulations. \pa{In order to circumvent this issue, one may assume  a short wavelength cut-off at the length scale of the monomer which would lead to a large weave-mode cut off and hence to a finite values of the sum in Eq.~\eqref{eq:denom}. However, this would imply that the longer the polymer the weaker the effect of the active force which is at odd with the numerical simulations (see Ref.~\cite{Bianco2018}).}


\subsection{Extended Rouse Model: introducing a finite bending rigidity}
\pa{In order to overcome  the diverging sum in Eq.~\eqref{eq:denom}} we propose to add a finite bending rigidity\pa{, $W$}
\begin{equation}
\dot{\mathbf{r}}(s,t)=-\mu W\partial_{s}^{4}\mathbf{r}(s,t)+
\mu Q\partial_{s}^{2}\mathbf{r}(s,t)+\mu f_0 \frac{\partial_{s}\mathbf{r}(s,t)}{|\partial_{s}\mathbf{r}(s,t)|}+\boldsymbol{\eta}(s,t)
\label{eq:Rouse-bend}
\end{equation}
\pa{since it is well known~\cite{Doi_book}, that by adding a finite bending rigidity the decay of the correlations between modes become 
\begin{align}
    \langle \mathbf{r}_n \mathbf{r}^*_n\rangle_0 \simeq \begin{cases}
    \frac{1}{n^2} & n \ll \frac{\ell_p}{b}\\
    \frac{1}{n^4} & n \gg \frac{\ell_p}{b}
    \end{cases}
\end{align}
and hence the sum in Eq.~\eqref{eq:denom} will keep finite even when $n\rightarrow \infty$.}
Notice that the ratio of the bending rigidity, $W$, and the strength of the monomer-monomer interactions  $Q$ identifies the persistence length
\begin{equation}
    \ell_p=b\sqrt{\frac{W}{Q}}
    \label{eq:lp}
\end{equation}
(where $b$ is the monomer length) and the number of Kuhn segments~\cite{Rubinstein} 
\begin{equation}
    N_p=\frac{b N}{\ell_p}
    \label{eq:Np}
\end{equation}
\pa{The advantage of assuming a finite bending rigidity instead of truncating the sum of the simple Rouse model in Eq.~\eqref{eq:Rouse} relies on the fact that numerical simulations dealing with self-avoiding polymers indeed introduce already a finite bending rigidity due to the excluded volume interaction between next-near-neighbors. Since this bending rigidity leads to a relatively short persistence length, in the following we will focus on the case in which the contribution of the bending rigidity is subdominant (a part of "regularizing" the sum in Eq.~\eqref{eq:denom}).}
The boundary conditions in the case fo a force- and torque-free polymer read~\cite{Winkler1995,Winkler2016}
\begin{subequations}\label{eq:BC_model}
\begin{align}
 \left[2Q\partial_s\mathbf{r}(s,t)-W\partial^3_s\mathbf{r}(s,t)\right]_{s=\pm N/2}&=0 \\
 \left[\sqrt{2QW}\partial_s\mathbf{r}(s,t)\pm W\partial^3_s\mathbf{r}(s,t)\right]_{s=\pm N/2}&=0
\end{align}
\end{subequations}
Also in this model, as in the previous one, we  avoid imposing the boundary conditions Eqs.~(\ref{eq:BC_model}) and, again, this amounts to introducing forces and torques on the edges of the polymer whose magnitude, direction and time correlation are, in principle, out of control and will be determined a posteriori.
In the Fourier representation, Eq.~\eqref{eq:Rouse-bend} becomes
\begin{align}
\sum_{n=-\infty}^{\infty}\dot{\mathbf{r}}_{n}(t)e^{\imath k_{n}s}=&-\mu \sum_{n=-\infty}^{\infty}k_{n}^{2}\left(Q+W k_{n}^{2}\right)\mathbf{r}_{n}(t)e^{\imath k_{n}s}\nonumber\\
&+\dfrac{\sum_{n=-\infty}^{\infty}   \imath \mu f_0 k_{n}\mathbf{r}_{n}(t)e^{\imath k_{n}s}}{\sqrt{\sum_{m=-\infty}^{\infty} k_m \mathbf{r}_{m}e^{\imath k_{m}s} \cdot\sum_{n=-\infty}^{\infty} k_n \mathbf{r}^*_{n}e^{-\imath k_{n}s}}}+\sum_{n=-\infty}^{\infty}\boldsymbol{\eta}_{n}(t)e^{\imath k_{n}s}
\end{align}
by multiplying both sides by $e^{-\imath k_{j}s}/N$, using  {the sum rule} (reported in Eq.~\eqref{eq:sum_rule}) and integrating in $ds$ we obtain:
\begin{align}\label{eq:mode_NL}
\sum_n &\frac{\sin\left(\frac{N}{2}(k_j-k_n)\right)}{N(k_j-k_n)}\dot{\mathbf{r}}_{\pa{n}}(t)=-\mu \sum_n k_{n}^{2}\left(Q+W k_{n}^{2}\right)\mathbf{r}_{n}(t)\frac{\sin\left(\frac{N}{2}(k_j-k_n)\right)}{N(k_j-k_n)}\nonumber\\
&+\frac{\imath \mu f_0}{N}\int \dfrac{\sum_{n=-\infty}^{\infty}  k_{n}\mathbf{r}_{n}(t)e^{\imath k_{n}s}}{\sqrt{\sum_{m=-\infty}^{\infty}k_m \mathbf{r}_{m}e^{\imath k_{m}s}\cdot \sum_{n=-\infty}^{\infty} k_n \mathbf{r}^*_{n}e^{-\imath k_{n}s}}} e^{-\imath k_{j}s} ds+\boldsymbol{\eta}_{j}(t)
\end{align}
\pa{As we did for the simple Rouse model, we  express the boundary conditions, Eq.~\eqref{eq:BC_model}, in their Fourier representation
\begin{subequations}\label{eq:BC_model_Fourier}
\begin{align}
 \sum_n \left[2Qk_n\mathbf{r}_n(t)+W k_n^3\mathbf{r}_n(t)\right]e^{\pm\imath k_n N/2} &=0 \\
 \sum_n \left[\sqrt{2 QW}k_n\mathbf{r}_n(t)\mp W k_n\mathbf{r}_n(t)\right]e^{\pm\imath k_n N/2}&=0\,.
\end{align}
\end{subequations}
In the numerical solutions, the series are truncated and these conditions are imposing two constraints on the $2N+1$ amplitudes of the Fourier modes. We assume that these constraints can be "absorbed" by the truncation procedure and we will check a posteriori that the effective forces and torques that we introduce with this ansatz are indeed disregardable.}
{One again we have to deal again with Eq.~\eqref{eq:denom}}: for this,  we now make the following \textit{ansatz} 
\begin{equation}\label{eq:apprx_ass}
   \underbrace{\sum_{n=-\infty}^{\infty}k_n^2 \mathbf{r}_{n}\cdot\mathbf{r}_n^*}_{\xi_0}\gg
   -\underbrace{\sum_{n=-\infty}^{\infty}\sum_{m\neq -n}k_m k_n \mathbf{r}_{m}\cdot\mathbf{r}_{n}e^{\imath (k_{m}+k_{n})s}}_{\xi} 
\end{equation}
whose validity  {will also be} checked \textit{a posteriori}.
Accordingly, Eq.(\ref{eq:mode_NL}) becomes:
\begin{align}\label{eq:mode_NL_red}
\sum_n\underbrace{ \frac{\sin\left(\frac{N}{2}(k_j-k_n)\right)}{N(k_j-k_n)}\dot{\mathbf{r}}_{\pa{n}}(t)}_{T_n}=&- \!\!\sum_{n=-\infty}^{\infty}\underbrace{\mu k_{n}^{2}\left(Q+W k_{n}^{2}\right)\mathbf{r}_{n}(t)\frac{\sin\left(\frac{N}{2}(k_j-k_n)\right)}{N(k_j-k_n)}}_{G_n}\nonumber\\
&+\!\!\sum_{n=-\infty}^{\infty}\underbrace{2\imath \mu f_0\dfrac{  k_n\mathbf{r}_{n}(t)\frac{\sin\left(\frac{N}{2}(k_j-k_n)\right)}{N(k_j-k_n)}}{\sqrt{\sum_{n=-\infty}^{\infty}k_n^2 \mathbf{r}_{n}(t)\cdot\mathbf{r}_n^*(t)}}}_{F_n}+\boldsymbol{\eta}_{j}(t)
\end{align}
{where we have introduced, for convenience, the functions $T_n$, $G_n$ and $F_n$.} It is interesting to notice that, at this stage, all terms contribute to the mixing of the modes,   
not only the non-linear terms $F_n$, as one would expect, due to the lack of orthogonality of the chosen basis. 

\noindent 
Before proceeding at numerically solving the model, we analyze the functions $T_n$, $G_n$ and $F_n$ in more detail. \pa{In order to do so, we focus on the behavior of these functions close to equilibrium for which we have 
\begin{align}
    \langle \mathbf{r}_n \mathbf{r}^*_n\rangle_0 \simeq \begin{cases}
    \frac{1}{n^2} & n \ll \frac{\ell_p}{b}\\
    \frac{1}{n^4} & n \gg \frac{\ell_p}{b}
    \end{cases}
\end{align} 
where $\ell_p$ is the persistence length (see Eq.~\eqref{eq:lp}).
In the following, we focus on (lower modes, $n\ll \ell/b$) modes for which the finite bending rigidity has not become relevant since the higher modes ($n\gtrsim \ell/b$) decay much faster. 
In particular, we look at the relative magnitude of the  terms with $n \neq j$ in Eq.~\eqref{eq:mode_NL_red}, as compared to  {the} mode $n=j$. }
\begin{figure}[h!]
    \centering
    \includegraphics[scale=0.28]{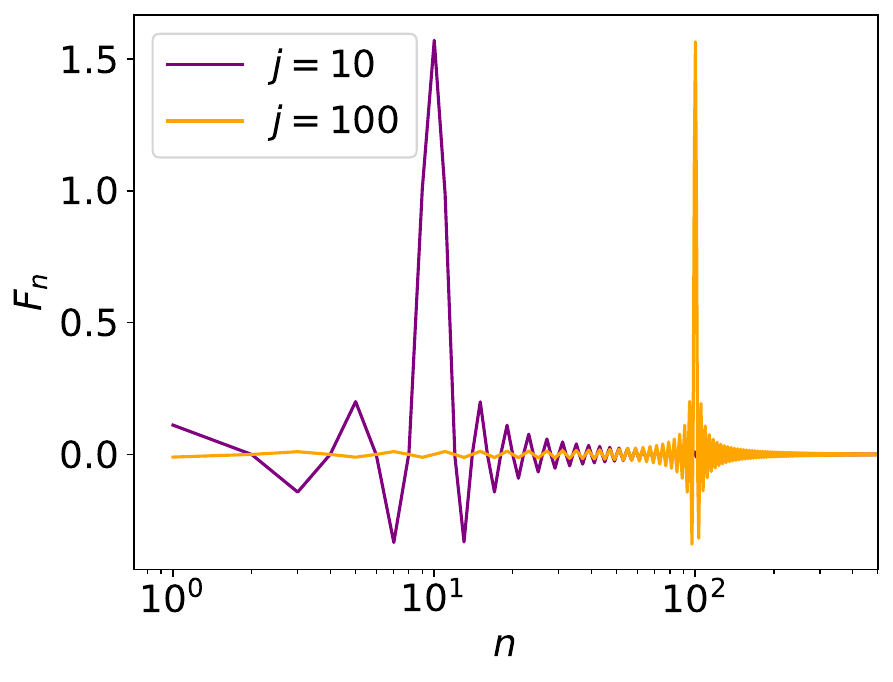}
    \includegraphics[scale=0.28]{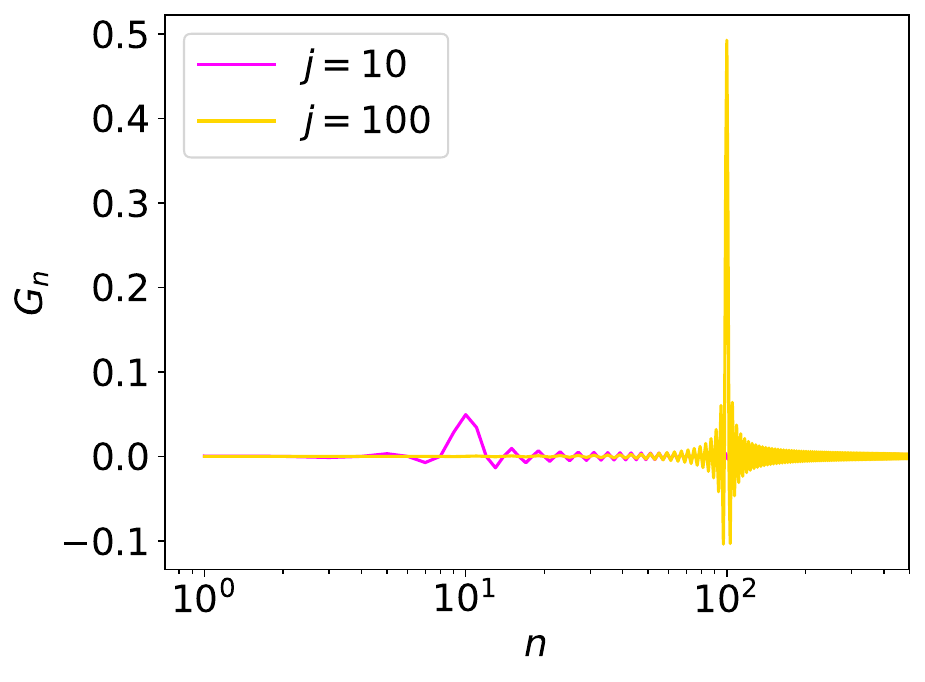}
    \includegraphics[scale=0.28]{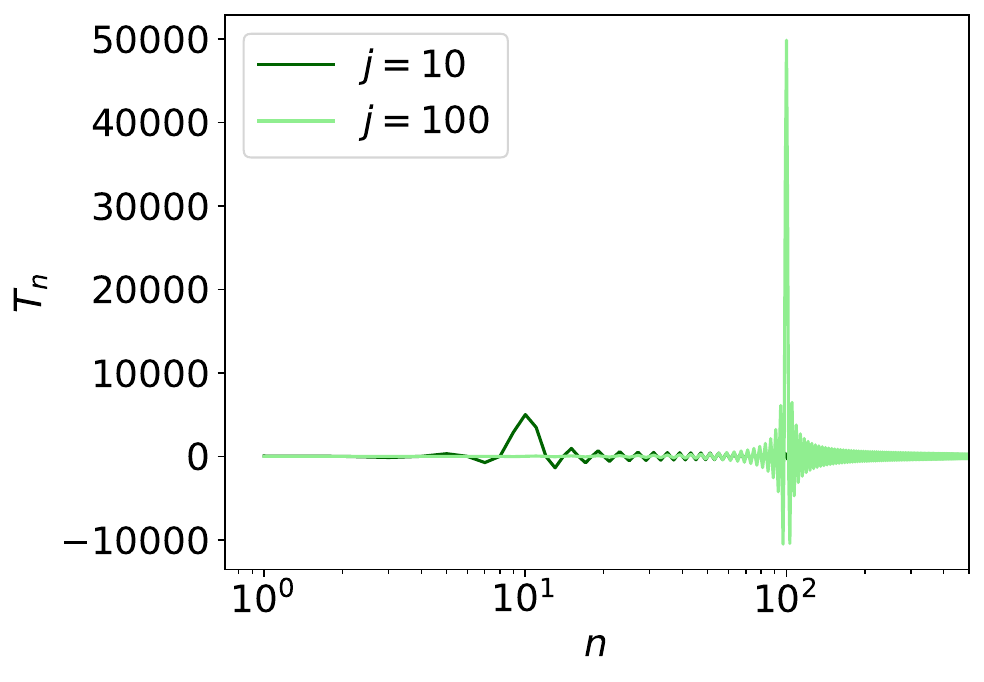}
    \caption{$F_n$ (left), $G_n$ (center), and $T_n$ (right) for $j=10,100$ \pa{where we assumed $\mathbf{r}_n\propto 1/n$ and $\dot{\mathbf{r}}_n\simeq \mathbf{r}_n/\tau_n\simeq n$ where we used $\tau_n \propto 1/n^2$.} We remark that $F_n$ contributions for $n\simeq 1$ are still relevant even for $j=100$. In contrast, for $T_n$ and $G_n$ once $j=100$ the modes with $n\simeq 1$ are disregardable.}
    \label{fig:est}
\end{figure}

\noindent
Figure~\ref{fig:est} shows that for $F_n$ several modes with $n\neq j$ still provide quite significant contribution to the overall sum as compared to $n=j$. \pa{In particular, many modes with $n\lesssim j$ provide significant contributions.} In contrast, for $T_n$ and $G_n$ the contributions from  the modes $n\neq j$ are much smaller and decay quickly, upon moving away from $n=j$. 
Accordingly, we approximate the sum involving $T_n$ and $G_n$ with the term with $n=j$ whereas we keep the full sum for $F_n$. \pa{This approach is not "exact" and its validity will be checked a posteriori by comparing it against the numerical simulations.}

\noindent
Hence we can rewrite Eq.~\eqref{eq:mode_NL_red} as 
\begin{equation}\label{eq:mode_NL_fin}
\dot{\mathbf{r}}_{j}(t)= - \mu k_{n}^{2}\left(Q+W k_{n}^{2}\right)\mathbf{r}_n(t)+\sum_{n=-\infty}^{\infty}\frac{2\imath \mu f_0 k_n \mathbf{r}_{n}(t)}{\sqrt{\sum_{n}k_n^2 \mathbf{r}_{n}(t)\cdot\mathbf{r}_n^*(t)}}\frac{\sin\left(\frac{N}{2}(k_j-k_n)\right)}{N(k_j-k_n)}+\boldsymbol{\eta}_{j}(t)
\end{equation}

Eq.~\eqref{eq:mode_NL_fin} represents our minimal model for the active polymer with bending. \pa{We remark that for $f_0 = 0$ our model reduces to the standard Rouse model with finite bending rigidity.} Finally, recalling that $k_j\equiv j\pi/N$ where $N$ is the number of monomers, we can rewrite Eq.~\eqref{eq:mode_NL_fin}
 as
\begin{equation}\label{eq:mode_NL_red_adim2}
\dot{\mathbf{r}}_{j}(t)=-\mu  n^{2}(Q_N+W_N n^2)+  \sum_{n=-\infty}^{\infty}\dfrac{\frac{2}{\pi}\imath\mu f_0 n \mathbf{r}_{n}(t)}{\sqrt{\sum_{n}n^2\mathbf{r}_{n}(t)\cdot\mathbf{r}_n^*(t)}}\frac{\sin\left(\frac{\pi}{2}(j-n)\right)}{j-n} +\boldsymbol{\eta}_{j}(t)
\end{equation}
We note that the last expression is independent of the polymer length $N$, since $N$ only appears in $Q_N = {Q\pi^2}/{N^2}$ and
$W_N = {W \pi^4}/{N^4}$ which, together with $f_0$ are the three parameters governing the dynamics of the system.  {In other words,} we can numerically solve (a finite set of) Eqs.~\eqref{eq:mode_NL_red_adim2} for different values of $Q_N, W_N$.
Given that 
\begin{align}
    \ell_p = b \sqrt{\frac{W}{Q}}=\frac{N b}{\pi}\sqrt{\frac{W_N}{Q_N}}
\end{align}
and using $\ell_p N_p = N b$ leads to
\begin{align}
    N_p = \pi\sqrt{\frac{Q_N}{W_N}}
\end{align}
Accordingly, the ratio $Q_N/W_N$, is proportional to the number of Kuhn segments.  

\noindent
We numerically integrate these equations    using the Euler algorithm, and compute the gyration radius  from the steady state value of the correlations among the modes (see Appendix A).



\section{Comparison of the gyration ratio predicted in the theory with the gyration ratio from computer simulations}
Using the definition of the Kuhn length,  {the equilibrium expression of the gyration radius for a Gaussian polymer} reads~\cite{Doi_book}:
\begin{equation}
    \label{eq:RG-doi}
    R_G=\ell_p \sqrt{\frac{N_p}{6}}
\end{equation}
{We aim to compare the predictions of the minimal model Eq.~\eqref{eq:mode_NL_fin} with the results of numerical simulations, reported in the literature\cite{bianco2018globulelike}. In this context,} it is convenient to express the gyration radius (whose dependence on the amplitude of the modes is reported in the Appendix A) in terms of $Q_N$ and $W_N$
\begin{equation}
    R_G=\sqrt{\frac{\pi k_BT}{2 Q_N}}\left(\frac{W_N}{Q_N}\right)^\frac{1}{4}
\end{equation}
where we used Eqs.~(\ref{eq:lp}), (\ref{eq:Np}) and $Q=3k_BT/b^2$, as in Ref.~\cite{Doi_book}. 
{Since, as mentioned, Eq.~\eqref{eq:mode_NL_fin} depends only on $f_0$, $W_N$ and $Q_N$, it is tempting to cast the model predictions, as well as the numerical data, in terms of these quantities.} Indeed, we find that  
for the extended Rouse model Eq.~\eqref{eq:mode_NL_fin},  {the ratio between the gyration radius of an active polymer and its equilibrium value, $R_G(N,\Pe)/R_G(N,0)$, collapse onto a master curve if plotted against the following quantity }
\begin{equation}
    \Pe_R=\frac{f_0 R_G}{k_BT}= \frac{f_0 \ell_p}{k_BT} \sqrt{\frac{N_p}{6}} =\frac{f_0 b}{k_BT}\sqrt{\frac{\pi}{2}\frac{k_BT}{b^2 Q_N}}\left(\frac{W_N}{Q_N}\right)^\frac{1}{4} 
    \label{eq:Pe_def}
\end{equation}
 {that is indeed a function of $f_0$, $W_N$ and $Q_N$; $b$ is the monomer size.}
 Eq.~\eqref{eq:Pe_def} shows that  polymers with a longer persistence length, characterized by a larger value of $R_G$, will  {be also characterized, at fixed $f_0$, by} a larger value of $\Pe_R$ (see also Appendix~\ref{app:f0}).

\begin{figure}[h!]
    \centering
     \includegraphics[scale=0.5]{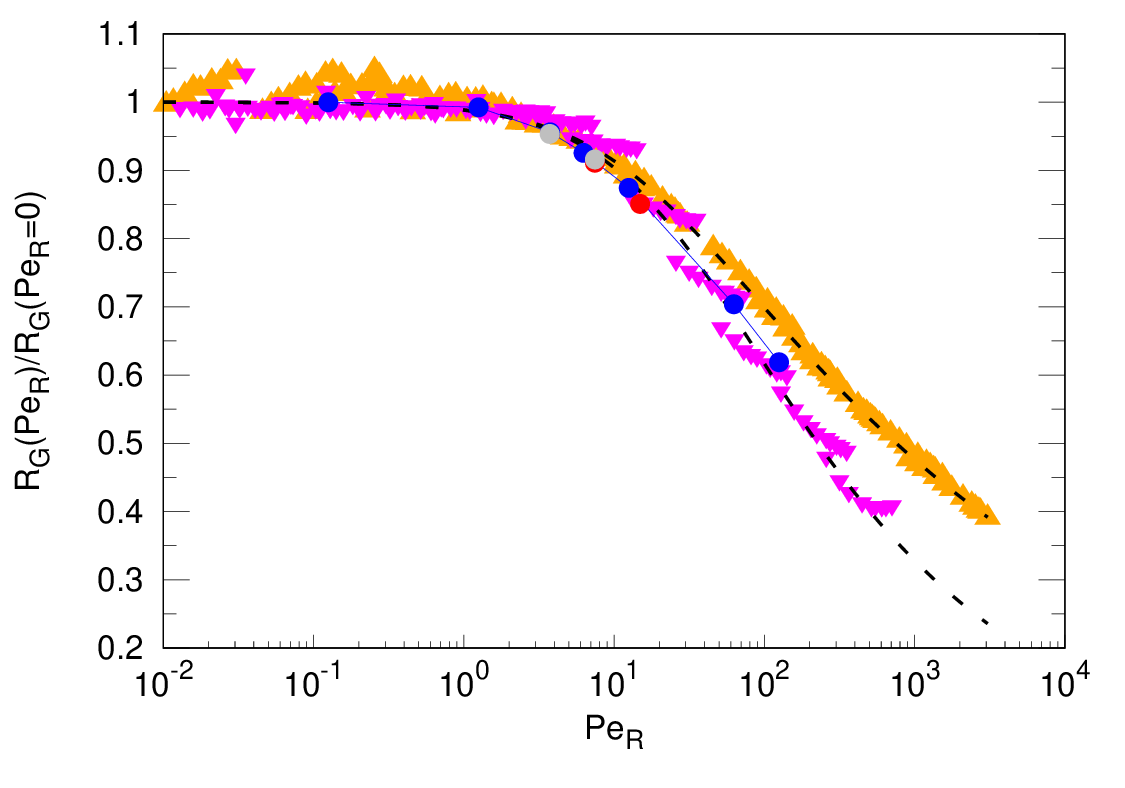}
     \caption{$R_G$ normalized by its equilibrium value as a function of $\Pe_R$ Eq.~\eqref{eq:Pe_def}. Triangles are the data from \cite{bianco2018globulelike}: orange upward for self-avoiding polymers and pink downwards for Gaussian polymers. Circles are predictions of the Rouse model \textit{with} bending rigidity Eq.~\eqref{eq:mode_NL_fin} with $Q=0.1$ and $W=0.001$ (blue), $Q=0.1$ and $W=0.002$ (red) and $Q=0.2$ and $W=0.001$ (grey). 
     The dashed lines are a fit to the Gaussian polymer data (pink triangles) with the fitting function Eq.~\eqref{eq:scal_gauss} \pa{and to the SAW polymer data (orange triangles) with the fitting function Eq.~\eqref{eq:scal_SAW}}. 
     }
     \label{fig:data_PRL_1}
\end{figure}

 \noindent
 Fig.\ref{fig:data_PRL_1}  {reports the model predictions (blue circles) as well as data from simulations of Gaussian (pink downward triangles) and self-avoiding (orange upward triangles) active polymers}.

\noindent
 {As shown,  plotting $R_G(N,\Pe)/R_G(N,0)$ as a function of $\Pe_R$ leads to a collapse of all data.} In each curve, we detect a crossover from a plateau to a decay: in the first regime the gyration radius does not depend on the Peclet number, whereas in the second regime the gyration radius decreases with activity, as a signature of a coil-to-globule transition.  


\noindent 
Moreover, the data of Rouse model can be well fitted by the following curve (bottom dashed line in Fig.\ref{fig:data_PRL_1})
\begin{align}
R^{Rouse}_G(\Pe_R)=\left(1+\frac{\Pe_R}{15}\right)^{-0.21}\label{eq:scal_gauss}
\end{align}
whereas the data for the self-avoiding polymer are better fitted by the following curve (top dashed line in Fig.\ref{fig:data_PRL_1})
\begin{align}
R^{SAW}_G(\Pe_R)=\left(1+\frac{\Pe_R}{15}\right)^{-0.176}\label{eq:scal_SAW}
\end{align}
The scaling functions in Eqs.~\eqref{eq:scal_gauss},\eqref{eq:scal_SAW} on the one hand confirm that the P\'eclet number defined in Eq.~\eqref{eq:Pe_def} is the proper dimensionless number capturing the collapse of the active polymers. On the other hand, they also capture that the value of $\Pe_R$ at which the crossover from plateau to collapse occurs is around $\Pe_R \simeq 15$ i.e. the ``advective'' term $f_0 R_G$  should indeed be quite larger then the thermal energy in order to be able to detect the collapse. 
More in detail the constraint $\Pe_R\gg 15$ can be read as 
\begin{align}
    \frac{f_0 b}{k_BT}\gg 15 \sqrt{\frac{6}{N}}\sqrt{\frac{b}{\ell_p}}
    \label{eq:cond}
\end{align}
which implies that longer $N\gg 1$ and stiffer $\ell_p\gg b$ polymers require a weaker value of $f_0$ to fulfill the condition in Eq.~\eqref{eq:cond}.

\begin{figure}[h!]
    \centering
    \includegraphics[width=0.45\textwidth]{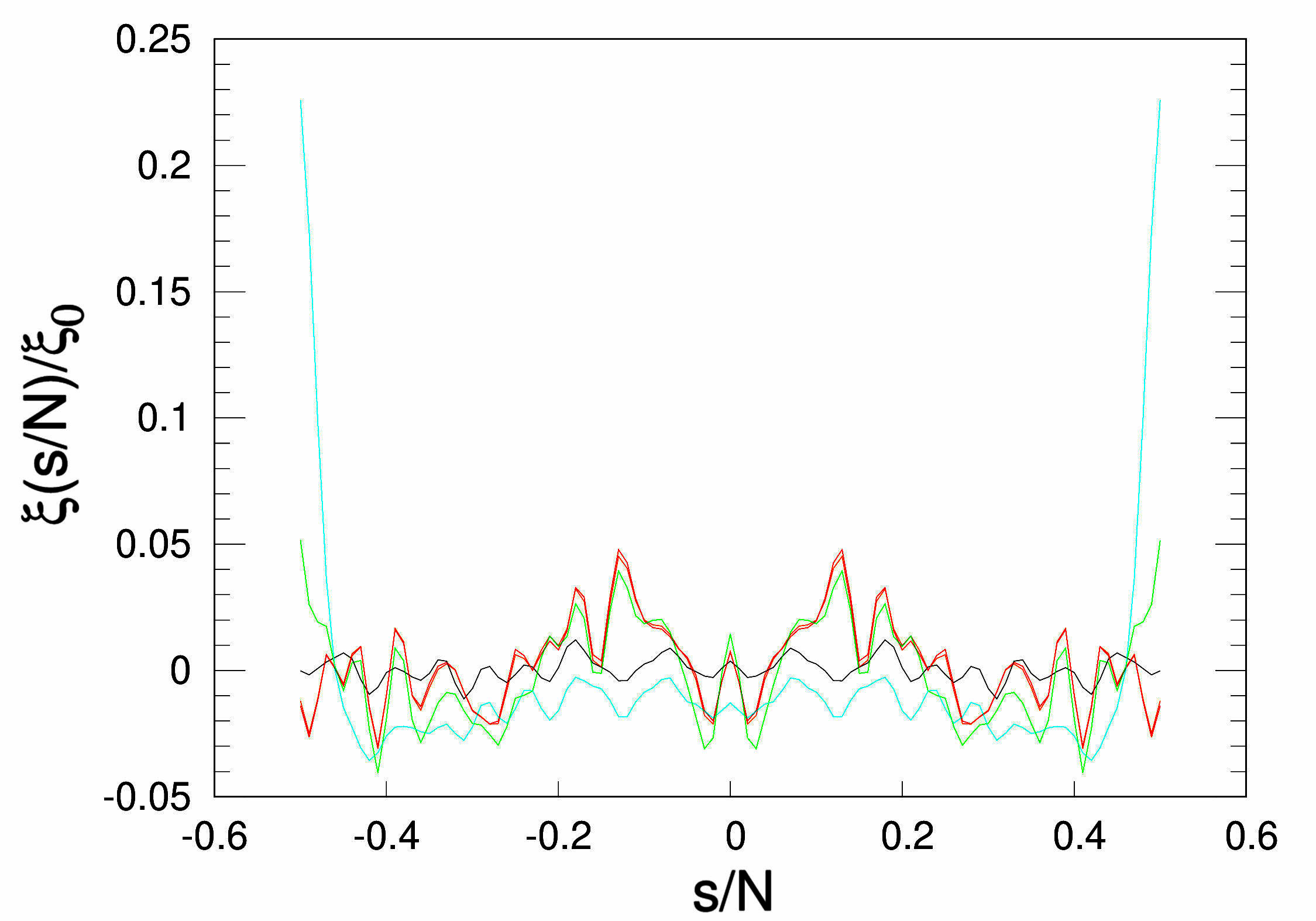}
    \includegraphics[width=0.45\textwidth]{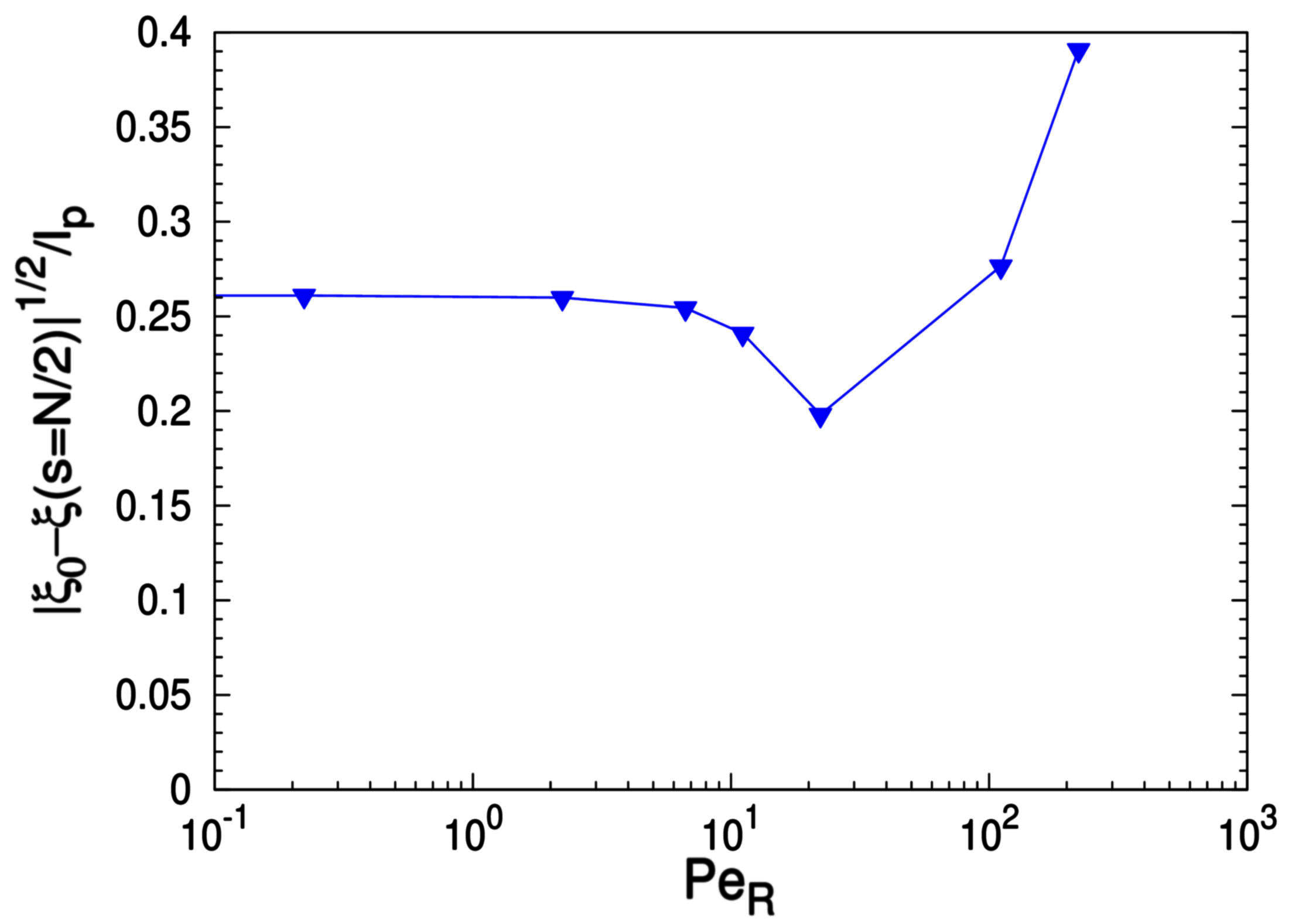}
    \caption{Check of the expansion assumption Eq.~\eqref{eq:apprx_ass}. Left panel: the value of the force is encoded in color code $f=0,0.1,1,5,10$ for black,red,green,cyan and $\ell_p/b = 0.1$  
    Right panel: $\xi/\xi_0(s=\pm N/2)$ as a function of the external force. 
    }
    \label{fig:check}
\end{figure}

\noindent
Finally, our model has been derived by means of some approximations. In particular the ansatz in Eq.~\eqref{eq:apprx_ass}  {needs to be verified} \textit{a posteriori}.  
{We report such verification in} Fig.~\ref{fig:check}. 

\noindent
 {In the left panel of Fig.~\ref{fig:check} we show the test of the assumption as a function of the normalised coordinate along the chain $s/N$ at different values of $f$. Indeed, the ratio $\xi/\xi_0$ is small everywhere for $f\lesssim 10$ and, for $f \geq 10$, it increases only in a small region close to the boundaries: overall the condition $\xi\ll \xi_0$ remains  satisfied.}  {Further, we verify, in the right panel of Fig.~\ref{fig:check}, that the net force that we apply at the ends is much smaller than the force applied at the middle of the backbone, even at large values of $\Pe_R$. }

\section{Conclusions}
In our work we have derived and analyzed an active Rouse model, extended to include finite bending rigidity, that retrieves one of the striking features of the tangentially active polymers with constant propulsion force: namely the collapse of the gyration radius upon increasing the active force on the backbone. The good quantitative agreement between the numerical integration of the active Rouse model and the simulations published in the literature allows us to derive few conclusions. First, our analysis leads us to identify $\Pe_R$ as the relevant dimensionless number that allows to collapse all  data onto a master curve. Second, once collapsed, the data can be fitted with a simple function that indeed may be employed to make predictions on the collapse of tangentially active polymers. Third, our model identifies the (normalized) tangential force as the term responsible for the correlations among the amplitudes of the Rouse modes. Such non-vanishing correlations indeed lead to the collapse of the gyration radius. Fourth, our model shows that in a pure active Rouse model, without bending rigidity and with an infinite number of modes, the active term would become vanishingly small due to the divergence of the denominator. On the one hand, one could set a cut off for the Rouse modes at some $n$ value. We show that, alternatively, a finite bending rigidity is sufficient in order to keep the active term finite.

\noindent
{However, it is also interesting to notice that the ``head-propulsion'', i.e. whether the free ends of the filament are self-propelled or not, has been shown to play a significant role on the steady-state conformations in case (i)\cite{li2023nonequilibrium, vatin2024conformation} where, for coherence, the ends are usually set as passive. On the contrary, in case (ii) and (iii) ends beads are always active. It would be interesting to elucidate the extent of the influence of this aspect, especially towards modelling real biophysical systems. Indeed, as worms and other filamentous organisms live in complex environments\cite{kudrolli2019burrowing}, it will be of capital importance to develop models able to capture the essential features of their locomotion, for robotic systems\cite{biswas2023dynamics} as well as for novel generation of filtering devices\cite{locatelli2023nonmonotonous}.}

\section*{Acknowledgements}
Our manuscript is part of the Special Issue in honour of Giovanni Ciccotti's birthday. We are very honoured to have had the possibility to meet him  and discuss  several scientific issues   with him. Even though we have never directly collaborated,    Giovanni has always been  a source of inspiration for all of us,  and for several generations of scientists. 

\noindent
E. Locatelli acknowledges support from the MIUR grant Rita Levi Montalcini. C.V acknowledges funding from MINECO grants C.V. acknowledges fundings   IHRC22/00002 and  PID2022-140407NB-C21 from MINECO.  

\section*{Data}
The data presented in this contribution can be found here: \href{10.5281/zenodo.10931600}{10.5281/zenodo.10931600}
\appendix

\section{Gyration tensor}
The gyration tensor is defined as:
\begin{equation}
    S_{ij}(t)=\frac{1}{bN}\int_{-\frac{N}{2}}^{\frac{N}{2}} (\mathbf{r}_i(s,t)-\mathbf{r}_{0,i}(t))(\mathbf{r}_j(s,t)-\mathbf{r}_{0,j}(t)) ds
\end{equation}
where $\mathbf{r}_0(t)$ is the location of the center of mass.
Exploiting the fact that $r$ is a real number, $\mathbf{r}_m^*=\mathbf{r}_{-m}$, then in the Fourier representation $S$ reads:
\begin{align}
    S_{ij}(t)=&\frac{1}{N^2}\int_{-\frac{N}{2}}^{\frac{N}{2}} \bigg[\sum_n \sum_m \mathbf{r}_{n,i}(t)\mathbf{r}^*_{m,j}(t)e^{\imath k_n s}e^{-\imath k_m s}- \mathbf{r}_{0,i}(t) \sum_n  \mathbf{r}^*_{n,j}(t)e^{-\imath k_n s}\nonumber\\
    &- \mathbf{r}^*_{0,j}(t) \sum_n  \mathbf{r}_{n,i}(t)e^{\imath k_n s} +\mathbf{r}_{0,i}\mathbf{r}^*_{0,j}\bigg]ds
\end{align}
that, using the sum rule in Eq.~\eqref{eq:sum_rule},  reads:
\begin{align}
    S_{ij}(t)&=\frac{1}{N^2}\int_{-\frac{N}{2}}^{\frac{N}{2}} \sum_n  \mathbf{r}_{n,i}(t)\mathbf{r}^*_{n,j}(t)-\mathbf{r}_{0,i}(t)\mathbf{r}^*_{0,j}(t)ds\nonumber\\
    &=\frac{1}{N}\sum_n  \mathbf{r}_{n,i}(t)\mathbf{r}^*_{n,j}(t)-\mathbf{r}_{0,i}(t)\mathbf{r}^*_{0,j}(t)\nonumber\\
    &=\frac{1}{N}\sum_{n\neq 0}  \mathbf{r}_{n,i}(t)\mathbf{r}^*_{n,j}(t) \label{eq:RG_sum}
\end{align}
Finally, the gyration radius is defined as $R^2_G=\text{Tr}S$ and, in equilibrium~\cite{Doi_book}, the amplitudes of the modes is given by Eq.~\ref{eq:equil}.
Hence the gyration radius reads
\begin{align}
    R_G^{eq} = b\sqrt{\frac{N}{6}}
\end{align}

\section{Sum Rule}
Coherently with Eq.~\eqref{eq:r_fourier} we define
\begin{align}
    \mathbf{r}_n(t)=\frac{1}{\sqrt{N}}\int_{-\frac{N}{2}}^{\frac{N}{2}}\mathbf{r}(s,t)e^{-\imath k_n s}ds\label{eq:sum_rule_1}
\end{align}
Substituting Eq.~\eqref{eq:r_fourier} into Eq.~\eqref{eq:sum_rule_1} we get:
\begin{align}
    \mathbf{r}_n(t)=\frac{1}{N}\int_{-\frac{N}{2}}^{\frac{N}{2}}\sum_{m=-\infty}^{\infty}\mathbf{r}_{m}(t)e^{\imath k_{m}s}e^{-\imath k_n s}ds
\end{align}
and using
\begin{align}
    \int_{-\frac{N}{2}}^{\frac{N}{2}}e^{\imath k_n s}e^{-\imath k_m s}ds=2\dfrac{\sin(\frac{N}{2}(k_m-k_n))}{k_m-k_n}
\end{align}
we get:
\begin{align}
    r_n=r_n+\frac{2}{N}\sum_{m\neq n}\dfrac{\sin(\frac{N}{2}(k_m-k_n))}{k_m-k_n}r_m
\end{align}
that leads to:
\begin{align}
    \frac{2}{N}\sum_{m\neq n}\dfrac{\sin(\frac{N}{2}(k_m-k_n))}{k_m-k_n}r_m=0
    \label{eq:sum_rule}
\end{align}

\section{Convergence of $R_G$ upon increasing the number of modes}
\begin{figure}[h!]
\centering
    \includegraphics[scale=0.45]{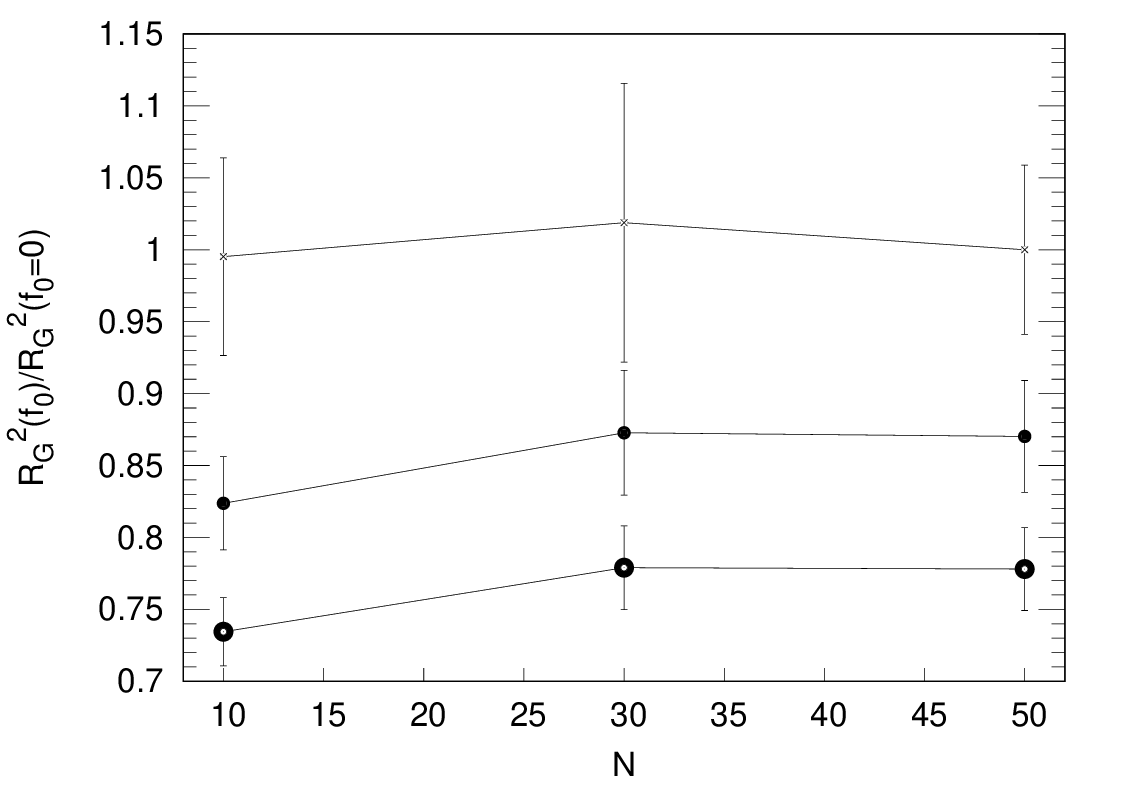}
    \caption{Convergence of $R_G$ upon increasing the number of modes (model with bending) for $\beta f_0 b=0,5,10$ and $\ell_p/b = 0.1$ standing bigger points for larger values of $f_0$.}
    \label{fig:RG_conv}
\end{figure}

Fig.~\ref{fig:RG_conv} shows the convergence of $R_G$ upon increasing the number of Rouse modes.  

\section{Effective force $\tilde{f}_0$}\label{app:f0}
It is interesting to notice that the denominator in the forcing term \pa{(see Eq.~\eqref{eq:mode_NL_red_adim2})} has the form 
\begin{align}
   \sum_{n=-\infty}^{\infty}n^2|\mathbf{r}_{n}(t)|^2 = 2\sum_{1}^{\infty}n^2|\mathbf{r}_{n}(t)|^2 
\end{align}
where we accounted for the fact that $|\mathbf{r}_0| < \infty$. For a semi-flexible polymer at equilibrium we have  
\begin{align}
    |\mathbf{r}_{n}(t)|^2\simeq \frac{1}{n^2\left(1+\frac{n^2}{n^2_p}\right)}
\end{align}
and hence the sum reads
\begin{align}
   \sum_{n=-\infty}^{\infty}n^2|\mathbf{r}_{n}(t)|^2 = 2\sum_{1}^{\infty} \frac{1}{1+\frac{n^2}{n^2_p}}\sim \frac{Nb}{\ell_p}
\end{align}
where we used $n_p = \frac{Nb}{\ell_p}$. This already provides an interesting scaling of the active force with the persistence length. Indeed the relevant parameter is 
\begin{align}
    \tilde{f}_0 \simeq f_0 \sqrt\frac{\ell_p}{Nb} 
\end{align}
Accordingly, using Eqs.~\eqref{eq:RG-doi},\eqref{eq:Np}, the P\'eclet number reads:
\begin{align}
  \Pe_R = \beta f_0 R_G = \beta \tilde{f}_0 \frac{bN}{\sqrt{6}}
\end{align}

\bibliographystyle{tfo}
\bibliography{active_polymer}

\begin{thebibliography}{76}
\providecommand{\url}[1]{\texttt{#1}}
\providecommand{\urlprefix}{URL }

\bibitem{marchetti2013hydrodynamics}
M.C. Marchetti, J.F. Joanny, S. Ramaswamy, T.B. Liverpool, J. Prost, M. Rao and
  R.A. Simha,  Reviews of modern physics  \textbf{85} (3), 1143 (2013).

\bibitem{fodor2018statistical}
{\'E}. Fodor and M.C. Marchetti,  Physica A: Statistical Mechanics and its
  Applications  \textbf{504}, 106--120 (2018).

\bibitem{cates2015motility}
M.E. Cates and J. Tailleur,  Annu. Rev. Condens. Matter Phys.  \textbf{6} (1),
  219--244 (2015).

\bibitem{schwarz2012phase}
J. Schwarz-Linek, C. Valeriani, A. Cacciuto, M. Cates, D. Marenduzzo, A.
  Morozov and W. Poon,  Proceedings of the National Academy of Sciences
  \textbf{109} (11), 4052--4057 (2012).

\bibitem{wittkowski2014scalar}
R. Wittkowski, A. Tiribocchi, J. Stenhammar, R.J. Allen, D. Marenduzzo and M.E.
  Cates,  Nature communications  \textbf{5} (1), 4351 (2014).

\bibitem{Palacci2013}
J. Palacci, S. Sacanna, A. Steinberg, D. Pine and P. Chaikin,  Science
  \textbf{339}, 936 (2013).

\bibitem{mognetti2013}
B. Mognetti, A. Sari\'c, S. Angioletti-Uberti, A. Cacciuto, C. Valeriani and D.
  Frenkel,  Physical Review Letters  \textbf{111}, 245702 (2013).

\bibitem{Cisneros2010}
L. Cisneros, R. Cortez, C. Dombrowski, R. Goldstein and J. Kessler,
  \emph{Animal Locomotion}, Vol. Springer pp. 99--115.

\bibitem{Murugan2015}
A. Murugan, J. Zou and M. Brenner,  Nat. communications  \textbf{6}, 1 (2015).

\bibitem{Mallory2018}
S. Mallory, C. Valeriani and A. Cacciuto,  Annual review of physical chemistry
  \textbf{69}, 59 (2018).

\bibitem{Narayan2007}
V. Narayan, S. Ramaswamy and N. Menon,  Science  \textbf{317}, 105 (2007).

\bibitem{Hayakawa2020}
Y. Hayakawa,  Europhysics Letters  \textbf{89}, 48004 (2010).

\bibitem{Suematsu2010}
N. Suematsu, S. Nakata, A. Awazu and H. Nishimori,  Physical Review E
  \textbf{81}, 056210 (2010).

\bibitem{fletcher2010cell}
D.A. Fletcher and R.D. Mullins,  Nature  \textbf{463} (7280), 485--492 (2010).

\bibitem{vale2003molecular}
R.D. Vale,  Cell  \textbf{112} (4), 467--480 (2003).

\bibitem{mahajan2022euchromatin}
A. Mahajan, W. Yan, A. Zidovska, D. Saintillan and M.J. Shelley,  Physical
  Review X  \textbf{12} (4), 041033 (2022).

\bibitem{goychuk2023polymer}
A. Goychuk, D. Kannan, A.K. Chakraborty and M. Kardar,  Proceedings of the
  National Academy of Sciences  \textbf{120} (20), e2221726120 (2023).

\bibitem{shin2024transcription}
S. Shin, G. Shi, H.W. Cho and D. Thirumalai,  Proceedings of the National
  Academy of Sciences  \textbf{121} (12), e2307309121 (2024).

\bibitem{loiseau2020active}
E. Loiseau, S. Gsell, A. Nommick, C. Jomard, D. Gras, P. Chanez, U. D’ortona,
  L. Kodjabachian, J. Favier and A. Viallat,  Nature Physics  \textbf{16} (11),
  1158--1164 (2020).

\bibitem{chakrabarti2022multiscale}
B. Chakrabarti, S. F{\"u}rthauer and M.J. Shelley,  Proceedings of the National
  Academy of Sciences  \textbf{119} (4), e2113539119 (2022).

\bibitem{chelakkot2014flagellar}
R. Chelakkot, A. Gopinath, L. Mahadevan and M.F. Hagan,  Journal of The Royal
  Society Interface  \textbf{11} (92), 20130884 (2014).

\bibitem{faluweki2023active}
M.K. Faluweki, J. Cammann, M.G. Mazza and L. Goehring,  Physical Review Letters
   \textbf{131} (15), 158303 (2023).

\bibitem{patra2022collective}
P. Patra, K. Beyer, A. Jaiswal, A. Battista, K. Rohr, F. Frischknecht and U.S.
  Schwarz,  Nature Physics  \textbf{18} (5), 586--594 (2022).

\bibitem{rosko2024cellular}
J. Rosko, K. Cremin, E. Locatelli, M. Coates, S.J. Duxbury, O.S. Soyer, K.
  Croft, K. Randall, C. Valeriani and M. Polin,  bioRxiv  pp. 2024--02 (2024).

\bibitem{deblais2023worm}
A. Deblais, K. Prathyusha, R. Sinaasappel, H. Tuazon, I. Tiwari, V.P. Patil and
  M.S. Bhamla,  Soft Matter  \textbf{19} (37), 7057--7069 (2023).

\bibitem{Dreyfus2005a}
R. Dreyfus, J. Baudry, M.L. Roper, M. Fermigier, H.A. Stone and J. Bibette,
  Nature  \textbf{437} (7060), 862--865 (2005).

\bibitem{Hill2014}
L.J. Hill, N.E. Richey, Y. Sung, P.T. Dirlam, J.J. Griebel, E. Lavoie-Higgins,
  I.B. Shim, N. Pinna, M.G. Willinger, W. Vogel, J.J. Benkoski, K. Char and J.
  Pyun,  ACS Nano  \textbf{8} (4), 3272--3284 (2014).

\bibitem{Biswas2017}
B. Biswas, R.K. Manna, A. Laskar, S. {Kumar P. B.}, R. Adhikari and G.
  Kumaraswamy,  ACS Nano  p. 10025 (2017).

\bibitem{Nishiguchi2018}
D. Nishiguchi, J. Iwasawa, H.R. Jiang and M. Sano,  New Journal of Physics
  \textbf{20} (1), 015002 (2018).

\bibitem{kumar2023emergent}
M. Kumar, A. Murali, A.G. Subramaniam, R. Singh and S. Thutupalli,  arXiv
  preprint arXiv:2303.10742   (2023).

\bibitem{subramaniam2024emergent}
A.G. Subramaniam, M. Kumar, S. Thutupalli and R. Singh,  arXiv preprint
  arXiv:2401.14178   (2024).

\bibitem{ozkan2021collective}
Y. Ozkan-Aydin, D.I. Goldman and M.S. Bhamla,  Proceedings of the National
  Academy of Sciences  \textbf{118} (6), e2010542118 (2021).

\bibitem{savoie2023amorphous}
W. Savoie, H. Tuazon, I. Tiwari, M.S. Bhamla and D.I. Goldman,  Soft Matter
  \textbf{19} (10), 1952--1965 (2023).

\bibitem{becker2022active}
K. Becker, C. Teeple, N. Charles, Y. Jung, D. Baum, J.C. Weaver, L. Mahadevan
  and R. Wood,  Proceedings of the National Academy of Sciences  \textbf{119}
  (42), e2209819119 (2022).

\bibitem{Winkler2020}
R.G. Winkler and G. Gompper,  The Journal of Chemical Physics  \textbf{153}
  (4), 040901 (2020).

\bibitem{ganai2014}
N. Ganai, S. Sengupta and G.I. Menon,  Nucleic acids research  \textbf{42} (7),
  4145--4159 (2014).

\bibitem{Smrek2017}
J. Smrek and K. Kremer,  Physical Review Letters  \textbf{118} (9), 1--5
  (2017).

\bibitem{Active_topoglass_NatComm20}
J. Smrek, I. Chubak, C.N. Likos and K. Kremer,  Nat. Commun.  \textbf{11} (1),
  26 (2020).

\bibitem{osmanovic2017dynamics}
D. Osmanovi{\'c} and Y. Rabin,  Soft matter  \textbf{13} (5), 963--968 (2017).

\bibitem{Kaiser2014}
A. Kaiser and H. L{\"o}wen,  The Journal of chemical physics  \textbf{141} (4)
  (2014).

\bibitem{eisenstecken2016conformational}
T. Eisenstecken, G. Gompper and R.G. Winkler,  Polymers  \textbf{8} (8), 304
  (2016).

\bibitem{Das2019}
S. Das and A. Cacciuto,  Physical Review Letters  \textbf{123} (8), 087802
  (2019).

\bibitem{theeyancheri2024dynamic}
L. Theeyancheri, S. Chaki, T. Bhattacharjee and R. Chakrabarti,  Physical
  Review Research  \textbf{6} (1), L012038 (2024).

\bibitem{prathyusha2022emergent}
K. Prathyusha, F. Ziebert and R. Golestanian,  Soft Matter  \textbf{18} (15),
  2928--2935 (2022).

\bibitem{Isele-Holder2015}
R.E. Isele-Holder, J. Elgeti and G. Gompper,  Soft Matter  \textbf{11} (36),
  7181--7190 (2015).

\bibitem{bianco2018globulelike}
V. Bianco, E. Locatelli and P. Malgaretti,  Physical Review Letters
  \textbf{121} (21), 217802 (2018).

\bibitem{Terakawa2017}
T. Terakawa, S. Bisht, J.M. Eeftens, C. Dekker, C.H. Haering and E.C. Greene,
  Science  \textbf{358} (6363), 672--676 (2017).

\bibitem{vliegenthart2020filamentous}
G.A. Vliegenthart, A. Ravichandran, M. Ripoll, T. Auth and G. Gompper,  Science
  advances  \textbf{6} (30), eaaw9975 (2020).

\bibitem{Grosberg2023}
I. Eshghi, A. Zidovska and A.Y. Grosberg,  Phys. Rev. Lett.  \textbf{131},
  048401 (2023).

\bibitem{nguyen2021emergent}
C. Nguyen, Y. Ozkan-Aydin, H. Tuazon, D.I. Goldman, M.S. Bhamla and O. Peleg,
  Frontiers in Physics  \textbf{9}, 734499 (2021).

\bibitem{mokhtari2019dynamics}
Z. Mokhtari and A. Zippelius,  Physical review letters  \textbf{123} (2),
  028001 (2019).

\bibitem{das2019dynamics}
S. Das and A. Cacciuto,  The Journal of Chemical Physics  \textbf{151} (24)
  (2019).

\bibitem{foglino2019}
M. Foglino, E. Locatelli, C. Brackley, D. Michieletto, C. Likos and D.
  Marenduzzo,  Soft matter  \textbf{15} (29), 5995--6005 (2019).

\bibitem{Locatelli2021}
E. Locatelli, V. Bianco and P. Malgaretti,  Phys. Rev. Lett.  \textbf{126},
  097801 (2021).

\bibitem{li2023nonequilibrium}
J.X. Li, S. Wu, L.L. Hao, Q.L. Lei and Y.Q. Ma,  Physical Review Research
  \textbf{5} (4), 043064 (2023).

\bibitem{miranda2023self}
J.P. Miranda, E. Locatelli and C. Valeriani,  Journal of Chemical Theory and
  Computation  \textbf{20} (4), 1636--1645 (2024).

\bibitem{vatin2024conformation}
M. Vatin, S. Kundu and E. Locatelli,  Soft Matter  \textbf{20} (8), 1892--1904
  (2024).

\bibitem{Lamura2024}
A. Lamura,  Phys. Rev. E  \textbf{109}, 054611 (2024).

\bibitem{Singh2018}
S.K. Anand and S.P. Singh,  Phys. Rev. E  \textbf{98}, 042501 (2018).

\bibitem{prathyusha2018dynamically}
K. Prathyusha, S. Henkes and R. Sknepnek,  Physical Review E  \textbf{97} (2),
  022606 (2018).

\bibitem{abbaspour2023effects}
L. Abbaspour, A. Malek, S. Karpitschka and S. Klumpp,  Physical Review Research
   \textbf{5} (1), 013171 (2023).

\bibitem{kurzthaler2021geometric}
C. Kurzthaler, S. Mandal, T. Bhattacharjee, H. L{\"o}wen, S.S. Datta and H.A.
  Stone,  Nature communications  \textbf{12} (1), 7088 (2021).

\bibitem{peterson2020statistical}
M.S. Peterson, M.F. Hagan and A. Baskaran,  Journal of Statistical Mechanics:
  Theory and Experiment  \textbf{2020} (1), 013216 (2020).

\bibitem{philipps2022tangentially}
C.A. Philipps, G. Gompper and R.G. Winkler,  The Journal of Chemical Physics
  \textbf{157} (19) (2022).

\bibitem{fazelzadeh2022effects}
M. Fazelzadeh, E. Irani, Z. Mokhtari and S. Jabbari-Farouji,  Phys. Rev. E
  \textbf{108}, 024606 (2023).

\bibitem{Farouji2023}
M. Fazelzadeh, Q. Di, E. Irani, Z. Mokhtari and S. Jabbari-Farouji,  The
  Journal of Chemical Physics  \textbf{159} (22), 224903 (2023).

\bibitem{Liebchen2019}
B. Liebchen and H. Löwen,  The Journal of Chemical Physics  \textbf{150},
  061102 (2019).

\bibitem{theurkauff2012dynamic}
I. Theurkauff, C. Cottin-Bizonne, J. Palacci, C. Ybert and L. Bocquet,
  Physical review letters  \textbf{108} (26), 268303 (2012).

\bibitem{Winkler1995}
L. Harnau, R.G. Winkler and P. Reineker,  The Journal of Chemical Physics
  \textbf{102} (19), 7750--7757 (1995).

\bibitem{Winkler2016}
T. Eisenstecken, G. Gompper and R.G. Winkler,  Polymers  \textbf{8} (2016).

\bibitem{Bianco2018}
V. Bianco, E. Locatelli and P. Malgaretti,  Phys. Rev. Lett.  \textbf{121},
  217802 (2018).

\bibitem{Doi_book}
M. Doi, S.F. Edwards and S.F. Edwards, \emph{The theory of polymer dynamics},
  Vol.~73   (, , 1988).

\bibitem{Rubinstein}
M. Rubinstein and R.H. Colby, \emph{Polymer Physics}   (Oxfrod university
  press, Oxford, 2003).

\bibitem{kudrolli2019burrowing}
A. Kudrolli and B. Ramirez,  Proceedings of the National Academy of Sciences
  \textbf{116} (51), 25569--25574 (2019).

\bibitem{biswas2023dynamics}
A. Biswas, T. Huynh, B. Desai, M. Moss and A. Kudrolli,  Physical Review Fluids
   \textbf{8} (9), 094304 (2023).

\bibitem{locatelli2023nonmonotonous}
E. Locatelli, V. Bianco, C. Valeriani and P. Malgaretti,  Physical Review
  Letters  \textbf{131} (4), 048101 (2023).

\end{thebibliography}

\end{document}